\documentclass[onecolumn,amsmath,superscriptaddress,notitlepage,nofootinbib,11pt]{revtex4-1}

\usepackage{pifont,enumerate}
\usepackage{graphicx,amsmath,amssymb}
\usepackage{dcolumn,mathtools}
\usepackage{bm}
\usepackage{times}
\usepackage{soul}
\usepackage[dvipsnames,table]{xcolor}
\definecolor{lightgray}{gray}{0.95}
\definecolor{darkgray}{gray}{0.8}

\usepackage{appendix}

\newcommand{\qed}{\nobreak \ifvmode \relax \else
      \ifdim\lastskip<1.5em \hskip-\lastskip
      \hskip1.5em plus0em minus0.5em \fi \nobreak
      \vrule height0.75em width0.5em depth0.25em\fi}

\begin{document}

\title{Universal Dynamics of Punctuated Progress}

\author{Yian Yin}
\affiliation{Department of Information Science, Cornell University, Ithaca, NY 14853, USA}
\affiliation{Center for Science of Science and Innovation, Northwestern University, Evanston, IL 60208, USA}
\affiliation{McCormick School of Engineering, Northwestern University, Evanston, IL 60208, USA}
\author{Dashun Wang}
\affiliation{Center for Science of Science and Innovation, Northwestern University, Evanston, IL 60208, USA}
\affiliation{McCormick School of Engineering, Northwestern University, Evanston, IL 60208, USA}
\affiliation{Kellogg School of Management, Northwestern University, Evanston, IL 60208, USA}
\affiliation{Ryan Institute on Complexity, Northwestern University, Evanston, IL 60208, USA}
\affiliation{Northwestern Innovation Institute, Northwestern University, Evanston, IL 60208, USA}

\begin{abstract}
Scientific and technological frontiers advance through punctuated dynamics~\cite{kuhn1997structure,gould1993punctuated,mokyr1990punctuated}, yet the principles governing these dynamics remain poorly understood.
Here we collect and analyze novel datasets tracking the evolution of frontiers across nine different domains, spanning materials discovery, structural biology, artificial intelligence, computational biomedicine, data science, theoretical computer science, Formula-1 racing, and physical wheel-building experiments. Analyzing over 6.8 million solutions to 6.7 thousand tasks, we uncover three universal patterns:
first, waiting times between new frontiers are heavy-tailed, with most attempts concentrated in long periods of stasis; second, frontier records accumulate at a sublinear rate, faster than logarithmic yet slower than linear growth; third, record-breaking events are temporally correlated, generating short-term predictability yet long-term unpredictability.
Despite the differences in the scale, scope, and definition of the nine settings, the empirical patterns are remarkably consistent across the diverse domains we study, and are not captured by existing models from complex systems~\cite{barabasi2005origin,karsai2012universal,wiser2013long,kauffman1987towards,bak1993punctuated,newman2003modeling,sole1996extinction,tria2014dynamics,loreto2016dynamics}, record statistics~\cite{krug2007records,berdahl2017records,nevzorov1988records}, economics of innovation~\cite{kortum1997research,evenson1976stochastic}, and cultural evolution~\cite{henrich2004demography,kolodny2015evolution,rendell2010copy,powell2009late}.
We trace the missing ingredient to the distinction between radical and incremental innovations~\cite{miu2018innovation,fleming2001recombinant,march1991exploration,liu2021understanding,christensen2013innovator,henderson1990architectural}, and develop a minimal, analytically solvable model incorporating both radical resets that restructure what is achievable and incremental refinements that exploit and compound around the current frontier.
Despite having a single tunable parameter, the model reproduces all three empirical regularities. Remarkably, the leading-order predictions are parameter-independent, identifying a new universality class governing punctuated progress and yielding testable predictions about how openness and access to frontier solutions shape the pace of advance.
Overall, these results reveal universal dynamics governing punctuated progress and identify the interplay between radical resets and incremental refinements as the key driver of how scientific and technological frontiers advance.
\end{abstract}

\maketitle
\clearpage

The frontiers of science and technology often exhibit punctuated dynamics~\cite{kuhn1997structure,gould1993punctuated},
yet our understanding of the basic principles governing such punctuated patterns remains limited, highlighting a longstanding gap in our quantitative understanding of scientific and technological progress.
Punctuated advances represent a foundational concept across the sciences~\cite{gould1989punctuated,gould2009punctuated,bak1993punctuated,elena1996punctuated,pagel2006large,thurner2010schumpeterian,mokyr1990punctuated,anderson1990technological,baumgartner2009punctuated,atkinson2008languages,valverde2015punctuated,loch1999punctuated,gersick1991revolutionary,duran2024multiscale}, appearing in settings ranging from evolutionary biology's punctuated equilibrium~\cite{gould1989punctuated, elena1996punctuated,sole1997self} to Kuhn's paradigm shifts~\cite{kuhn1997structure,bettencourt2009scientific}, and from the evolution of language~\cite{atkinson2008languages,valverde2015punctuated} to policy cycles~\cite{baumgartner2009punctuated}, each underscoring the fundamental dynamics of episodic, abrupt, and sometimes profound shifts.
Understanding when and how new frontiers emerge would not only deepen our knowledge of scientific and technological progress, but also illuminate the conditions under which breakthroughs occur, stagnation persists, and the pace of advance can be accelerated or slowed.

Here we systematically collect and analyze comprehensive databases tracing the dynamics of frontiers across nine different domains of science and technology (See Table~\ref{Tab1}, SI S1).
$D_1 - D_4$ capture scientific discoveries in four crucial domains that are advancing at rapid rates, often driven by large global investments and intense competitive search.
$D_1$ follows advances in materials science, by tracking the critical temperature of superconducting materials ($T_c$), building on 38,576 records (19,951 unique materials) from Reaxys, grouped into 33 families by chemical formula.
$D_2$ measures the progress of protein structure determination in biology, by tracing the evolution of R-free, a canonical metric of model-data fit, across 143,343 entries submitted to the Protein Data Bank targeting 3,644 unique protein clusters.
$D_3$ traces the development of artificial intelligence using the paperswithcode data~\cite{martinez2021research,ott2022mapping}, which tracks the progress of 385 AI tasks and compares the performance of 10,439 AI algorithms designed to advance these tasks.
$D_4$ covers research-level competitions in computational biomedicine, tracking 70 DREAM Challenges (Dialogue on Reverse Engineering Assessment and Methods)~\cite{margolin2013systematic}, including a total of 15,529 solutions submitted.

In parallel, large-scale algorithmic search ecosystems, including data science challenges and computational optimization competitions, generate vast numbers of solutions under well-specified evaluation metrics. $D_5$ comes from Kaggle, one of the largest data science communities, tracing the progress of 2,067 Kaggle competitions, totaling 5,792,702 submissions. $D_6$ captures 414 data science matches on TopCoder, including 188,807 submissions. $D_7$ records 19 programming competitions in theoretical computer science~\cite{miu2018innovation}, solving NP-hard problems, with participants contributing a total of 45,914 algorithmic solutions.

Finally, we examine complex engineered systems.
$D_8$ traces Formula-1 racing, representing one of the frontier technologies in engineering, where we examine the improvement in speed dynamics from 589,081 records across 41 circuits from 1996 to 2024.
Going beyond observational data, $D_9$ captures experimental data from two micro-society lab studies on cumulative cultural evolution based on a physical wheel-building task~\cite{derex2019causal,osiurak2021technical}, with 280 participants organized into 56 chains of cultural transmission and generating 1,400 solutions in total.

\begin{figure*}[htbp]
    \centering
    \includegraphics[width=1\columnwidth]{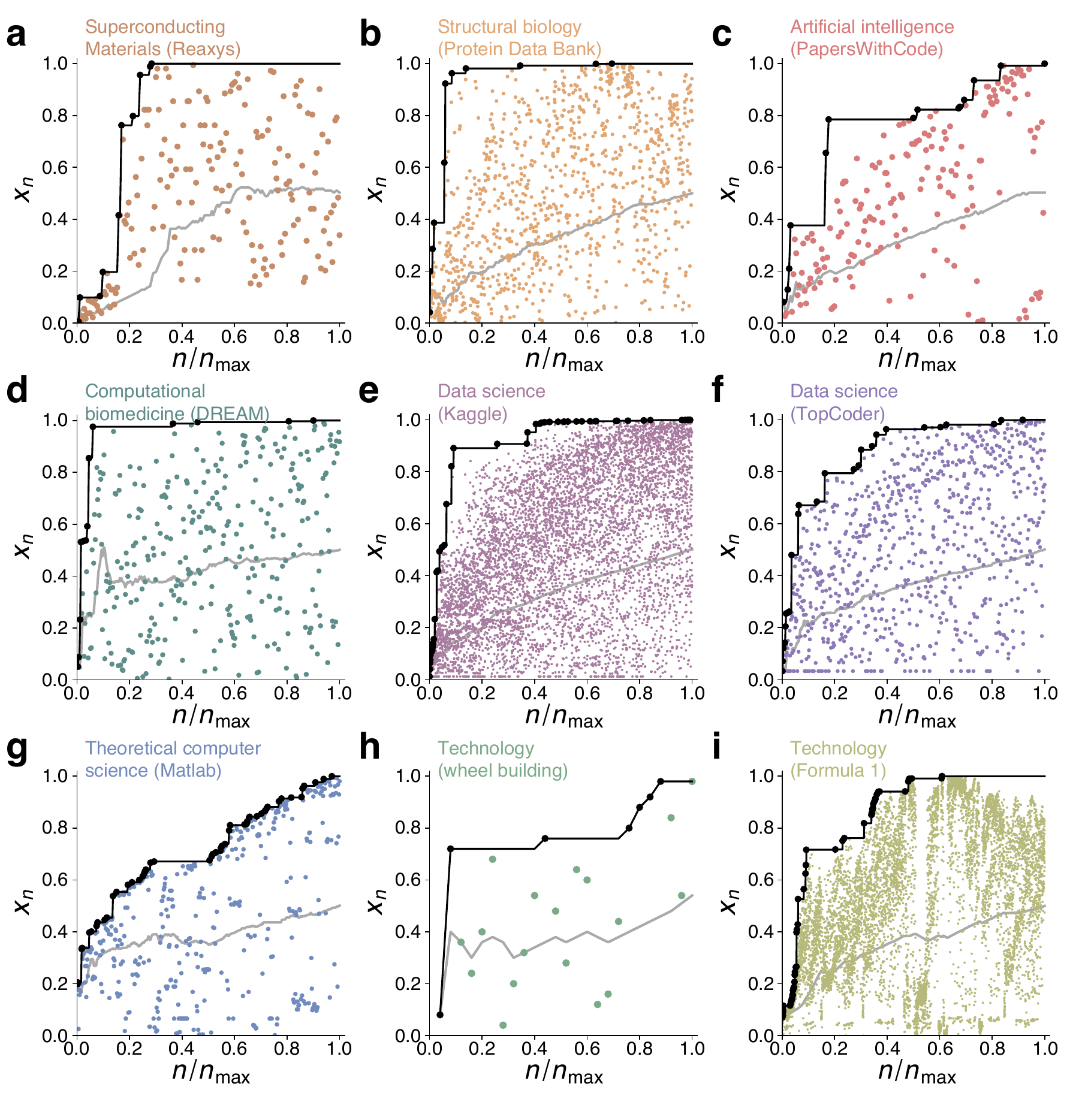}
  \caption{\textbf{Punctuated dynamics of science and technology frontiers.} \textbf{a-i}, Performance dynamics based on randomly selected examples from each of our nine datasets. Each black dot represents an individual solution, plotted by its relative temporal position ($n/n_{\max}$, x-axis) and performance ($x_n$, y-axis) within the sequence of submissions. The performance value $x_n$ is normalized as the percentile. The red curve highlights the frontier dynamics, showing the evolution of state-of-the-art performance over time. The grey curve indicates the median of $x_n$ up to a certain time point.}
  \label{fig:fig1}
\end{figure*}

Together, these nine datasets span scientific discovery (superconductors, structural biology, artificial intelligence, and computational biomedicine), algorithmic search (Kaggle, TopCoder, and NP-hard programming contests), and engineered systems (Formula 1 and wheel-building experiments), offering a broad view of how frontiers advance across modern science and technology (Table~\ref{Tab1}, Fig.~\ref{fig:fig1}). Despite their diversity, these systems share two key features. First, each comprises many distinct tasks or problem instances, together with the full sequence of solutions contributed over time. Second, every solution is evaluated using a well-defined performance metric, allowing us to rank all attempts and identify, at each step, whether a new record has been set.

To build intuition, we first highlight one task from each domain.
Figure~\ref{fig:fig1}a--i illustrate the punctuated frontier dynamics together with the numerous solutions beneath the frontier, depicting several intriguing patterns.
First, while the frontiers evolve in a punctuated manner, there is notable heterogeneity across these examples, with some showing rapid progressions in establishing new frontiers while others showing long periods of stasis.
Second, competitions tend to start with low performance,
but as they unfold, they show improvement in performance on average (grey line),
suggesting the performance distribution is not stationary in time.
Yet, despite the average performance improvements, the overall rate of improvement seems modest. Further, even for later stages of the competitions, low-quality solutions appear rather ubiquitous across different settings.
These examples suggest rich dynamics beneath the frontier, raising an interesting question: are there reproducible patterns governing the emergence of new frontiers?

To address this question, we formalize frontier dynamics within a single quantitative framework. Let $x_n$ denote the performance of the $n$-th attempt within a given task, and define the frontier value after $n$ attempts as
the record statistics $x^*_n \equiv \max \{x_1,\cdots,x_n\}$.
We call the $n$-th attempt record-breaking if it improves on the current frontier, and denote this with
$s_n\equiv\Theta (x^*_n-x^*_{n-1})=\Theta (x_n-\max \{x_1,\cdots,x_{n-1}\}),$
where $\Theta$ is the Heaviside step function and $s_1=1$.
The cumulative number of records established in the first $n$ attempts is then
$S_n\equiv \sum_{k=1}^{n}s_k$.
Inversely, the attempt at which the $s$-th record is set is $N_s\equiv \min\{n|S_n=s\}$ (Fig.~\ref{fig:fig2}a). Together, these quantities allow us to characterize three key aspects of frontier dynamics: how long systems wait between records, how quickly records accumulate, and how strongly past records shape future ones.

\section{Empirical regularities in frontier dynamics}
Across the nine frontier-setting systems we study, the dynamics of record-breaking events obey three robust regularities that hold despite large differences in domain, scale, and measurement.

\subsection{Regularity 1: Bursty waiting times and long stasis}

We first examine the waiting time between new frontiers, $W_n\equiv N_{S_n+1}-n$, which measures how many additional attempts are needed, after the $n$-th attempt, until the next record is set (Fig.~\ref{fig:fig2}a). Across all nine systems, from superconducting materials and structural biology to AI benchmarks, data science challenges, and Formula 1, we find that the distribution of waiting times is consistently heavy-tailed (Fig.~\ref{fig:fig2}b).
When rescaled by the mean waiting time,
$\langle W_n\rangle$,
the distributions collapse onto a common curve well approximated by a power-law tail:
\begin{equation}
   P(W_n=w)\sim \frac{1}{\langle W_n\rangle(w/\langle W_n\rangle)^{\gamma}},~~\gamma\in [2,3],
      \label{eq:data_dist}
\end{equation}

Note that there are well-known challenges in detecting power law distributions in empirical data~\cite{clauset2009power,voitalov2019scale,zhou2020power,serafino2021true}.
If $P(W_n)$ indeed follows a power law, one prediction is that the vast majority of the efforts are concentrated within a few long periods of stasis, akin to a Levy flight where the overall displacement is dominated by occasional long jumps~\cite{koren2007leapover,song2010modelling,zaburdaev2015levy}.
To test this, we measure the length of recent stasis,
defined as the number of attempts since the last record,
$Q_n\equiv n-N_{S_n}$.
In a non-power-law setting,
one would expect that the fraction $\frac{\langle Q_n\rangle}{n}$ converges to 0 for large $n$.
By contrast, a power-law tail
would predict a nearly constant fraction $\frac{\langle Q_n\rangle}{n}$ as a function of $n$, as $Q_n$ grows in proportion to the total number of attempts $n$.
We test this prediction across our nine datasets, finding that the data show systematic support for the power-law predictions (Fig.~S3).

Overall, this first regularity demonstrates that frontier dynamics is intrinsically bursty: most new records arrive in short bursts, separated by unusually long periods of stasis. In other words, the typical trajectory is not gradual progress with occasional pauses, but long plateaus punctuated by rare, clustered jumps.

\subsection{Regularity 2: Sublinear growth of frontier records}

We next ask how quickly the number of frontiers accumulates as attempts increase, denoting with
$S_n=\sum_{k=1}^n s_k$
the number of records achieved in the first $n$ attempts. Across all nine systems, we find that $\langle S_n\rangle$ consistently grows in a sublinear yet faster-than-logarithmic manner (Fig.~\ref{fig:fig2}c).

Indeed, the empirical patterns we observe lie strictly between two canonical baselines: (i) Logarithmic growth,  $\langle S_n\rangle\sim \ln n$, predicted by record statistics for independent draws from a fixed distribution~\cite{kortum1997research,evenson1976stochastic,terwiesch2008innovation,lemus2021dynamic,krug2007records,berdahl2017records,nevzorov1988records}, which predicts that breakthroughs become exponentially harder over time; (ii) $\langle S_n\rangle \sim n$, predicted by cumulative imitation models~\cite{henrich2004demography,kolodny2015evolution,rendell2010copy,powell2009late}, where each new attempt has a constant probability of surpassing the current best. In contrast, the data follow an intermediate regime: new frontiers arrive faster than random chance would allow, but slower than a world where each new attempt has a fixed chance of being a breakthrough.

This second regularity therefore shows that frontiers do not simply ``get harder'' in the standard record-statistics sense. Instead, record opportunities renew often enough to sustain progress well above logarithmic growth, yet not so easily as breakthroughs arrive at a constant rate.

\subsection{Regularity 3: Temporal correlations and amplified variance}

Finally, we investigate how past frontier events correlate with future ones.
A fundamental assumption of current modeling frameworks posits that record-breaking activities are memoryless~\cite{henrich2004demography,kortum1997research}. Hence, the indicator variables $s_n$ should be independent over time, and the time elapsed since the last record and the time until the next should be uncorrelated.
Yet empirically, we find the opposite. Across all systems we examine, there is a systematic positive correlation between ``time since last record'' and ``time to next record''. Comparing
$Q_n$, number of attempts since the last record, with $W_n$, number of attempts until the next record, we observe a consistent correlation between
$W_n/\langle W_n\rangle$ and $Q_n/\langle Q_n\rangle$
across domains (Fig.~\ref{fig:fig2}d).
This means the frontier evolution is path-dependent: the longer a system has gone without a record, the longer it is likely to wait for the next one. Conversely, periods with frequent records tend to be followed by further breakthroughs, showing an interesting path dependence, where recent history tilts the odds of near-future breakthroughs.

The uncovered temporal correlations have a striking consequence. If record-breaking events were uncorrelated, fluctuations in the cumulative number of records would obey a central-limit scaling, with the mean square fluctuation $F_n\equiv \langle (S_n-\langle S_n\rangle)^2\rangle =\langle S_n^2\rangle - \langle S_n\rangle^2$
bounded by the mean~\cite{sole1997self},  $F_n\leq \langle S_n\rangle$. Instead, we find that
the growth in $F_n$ is systematically higher than expected (Fig.~\ref{fig:fig2}e), approximately following $\langle S_n\rangle^2$ (Fig.~\ref{fig:fig2}e). In other words, the systems we examine violate the self-averaging assumptions behind standard forecasting: Because bursts breed bursts and stalls breed stalls, early fluctuations lock in long-run trajectories, resulting in long-run outcomes that are far more heterogeneous and unpredictable than existing models predict (SI S2). Some instances progress much faster than the average trajectory, others much slower, even when they belong to the same domain. Together, frontier dynamics exhibit both short-run predictability and long-run unpredictability, as fluctuation is amplified rather than averaged away.

Taken together, these three empirical regularities---bursty waiting times with long stasis, sublinear but faster-than-logarithmic growth of frontier records, and strong temporal correlations that amplify long-run variance---unveil a new quantitative picture of how scientific, algorithmic, and technological frontiers evolve (Fig.~\ref{fig:fig2}). No existing framework from complex systems~\cite{barabasi2005origin,karsai2012universal,wiser2013long,kauffman1987towards,bak1993punctuated,newman2003modeling,sole1996extinction,tria2014dynamics,loreto2016dynamics}, record statistics~\cite{krug2007records,berdahl2017records,nevzorov1988records}, economics of innovation~\cite{kortum1997research,evenson1976stochastic}, or cultural evolution~\cite{henrich2004demography,kolodny2015evolution,rendell2010copy,powell2009late},
simultaneously anticipates all three patterns (SI S2): random-search models~\cite{kortum1997research,evenson1976stochastic,terwiesch2008innovation,lemus2021dynamic,krug2007records,berdahl2017records,nevzorov1988records} capture heavy-tailed waiting times but predict too few records and no temporal correlation; cumulative learning models predict rapid record accumulation, with poissonian and memoryless dynamics~\cite{henrich2004demography,rendell2010copy,thompson2022complex}.

A common limitation of these models is that they treat all attempts as homogeneous: every new solution is drawn from the same distribution or generated by the same mechanism. Yet, across our domains, frontier advances are driven by a mix of qualitatively different moves: radical innovations that reconfigure solutions and open new regions of the performance landscape, and incremental improvements that refine existing approaches. This suggests that the missing ingredient may be the interaction between these two innovation modes. We therefore seek a minimal mechanistic model that distinguishes radical from incremental innovations and ask whether their interplay is sufficient to generate the punctuated dynamics we observe.

\section{Modeling record-breaking dynamics}

Here we develop a simple model in which each new attempt arises from one of two innovation modes: a radical innovation that proposes a wholly new solution, or an incremental innovation that modifies and improves components of the current frontier. This radical-incremental dichotomy has deep roots in the innovation literature, appearing as leaps vs.~tweaks~\cite{miu2018innovation}, global vs.~local searches~\cite{stuart1996local,fleming2001recombinant}, explorations vs.~exploitations~\cite{march1991exploration,liu2021understanding}, sustaining vs.~disruptive technologies~\cite{christensen2013innovator}, architectural vs.~modular innovations~\cite{henderson1990architectural}, and paradigm shift vs.~normal science~\cite{kuhn1997structure}. Yet it has not been explicitly incorporated into mathematical models for frontier dynamics. Next we show that doing so yields a one-parameter model that analytically reproduces all three empirical regularities.

Our model mimics how agents develop new solutions based on existing advances (Fig.~\ref{fig:fig3}). Each solution is viewed as a combination of $B\rightarrow\infty$ components~\cite{bak1993punctuated}, where each component is associated with a score $x^{(b)}$ for $1\leq b\leq B$. The overall performance of the solution is a weighted sum of these components, where the weight $w_{b}$ approximates the relative importance of component $b$. The overall performance score of the $n$-th attempt can, therefore, be formulated as
$$x_n=\sum_{b=1}^B w_{b}x_n^{(b)}.$$

Our model formalizes a minimal two-channel search mechanism, where an innovator can take either i) \emph{radical} resets with probability $p_r$, which restructure what is achievable, or ii) incremental refinements with probability $p_i=1-p_r$, which exploit the current frontier. In radical innovations, the innovator chooses to draw new random scores $x_{n}^{(b)}\sim U[0,1]$ for \emph{all} components $b=1,\cdots,B$, independent of previous versions~\cite{bednorz1986possible}.
By contrast, in incremental innovations, one builds on the state-of-the-art solution and improves each component one by one. Here the innovator focuses on one component $b_n$ each time, replacing this component with a new version (a random draw) while keeping all other components unchanged, until a new frontier $x_n^{(b)}>x_n^{(b)*}$ emerges:
\begin{equation}
x_n^{(b)}\sim I(b\neq b_n)\delta(x-x_{n}^{(b)*})+I(b= b_n)U[0,1],
\label{eq:bn_transition}
\end{equation}
where $x_{n}^{(b)*}\equiv\max\{x_{1}^{(b)},\cdots,x_{n-1}^{(b)}\}$ represents the best version to date for component $b$. After a successful improvement, the agent switches to the next component ($b\rightarrow b+1$) and repeats the process.
Hence, the evolution of component $b_n$ can be described as a stochastic process, following
\begin{equation}
b_{n+1}=b_{n}+\Theta(x_n^{b}-x_{n}^{(b)*})
\label{eq:an_bn_transition}
\end{equation}

For simplicity, we assume that the variance and uncertainty in radical innovations are larger than the (accumulated) differences from incremental innovations~\cite{romanelli1994organizational}. Overall, the model has one tunable parameter $p_r$, balancing between radical and incremental approaches, prompting us to call it the $p$ model.

\section{Model predictions}
\begin{figure*}[htbp]
    \centering
    \includegraphics[width=1\columnwidth]{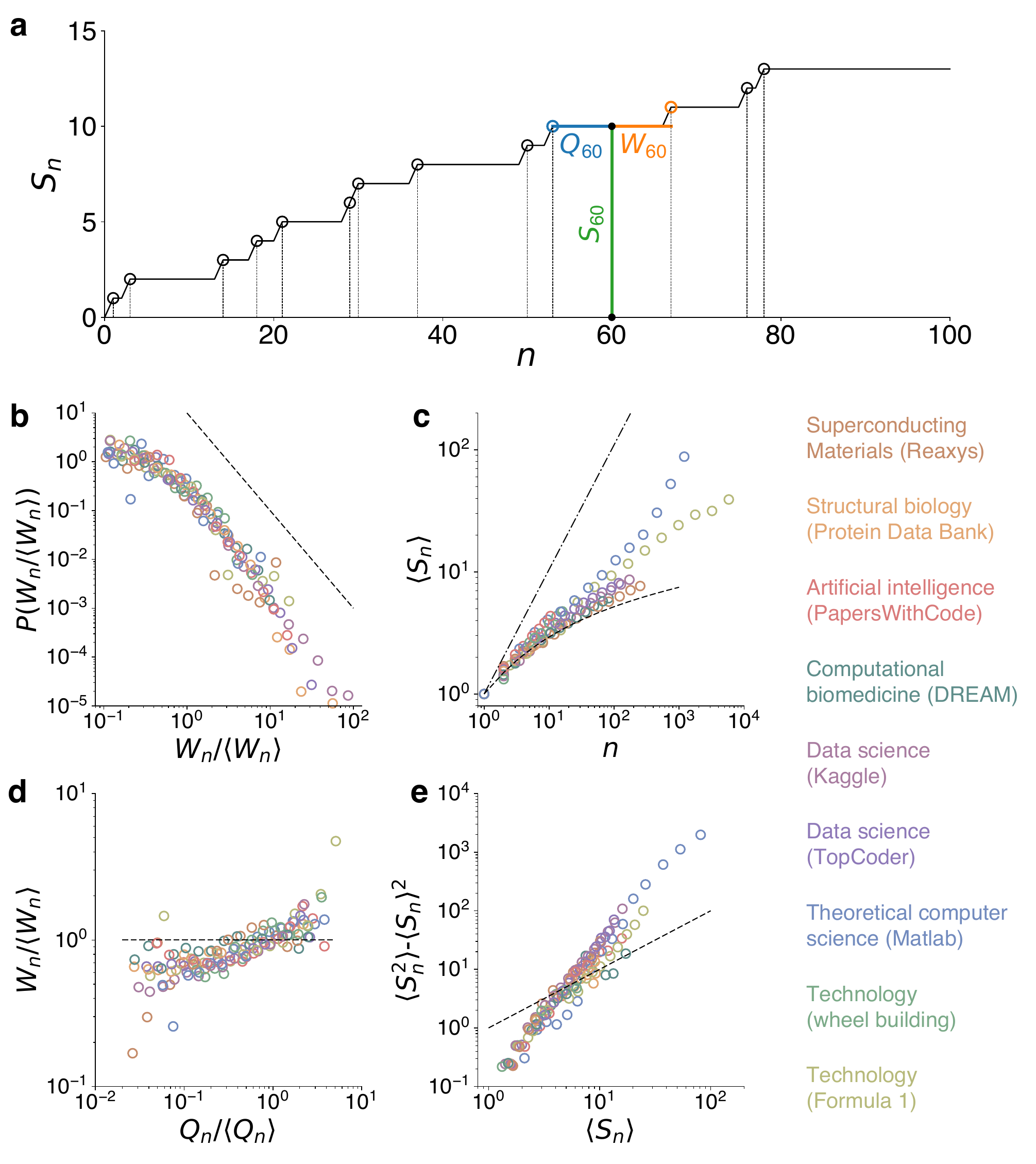}
  \caption{\textbf{Quantifying the punctuated frontier dynamics.}
\textbf{a}, Key quantities that characterize the frontier dynamics. $S_n$ denotes the number of record-breaking events in the first $n$ attempts. $W_n\equiv N_{S_n+1}-n$ measures the waiting time until the next frontier, $N_{S_n+1}$, while $Q_n\equiv n-N_{S_n}$ measures the time elapsed since the last frontier $N_{S_n}$.
\textbf{b}, The waiting time features a fat-tailed distribution, approximately following a power-law tail $P(W_n)\sim W_n^{-\gamma}$. Dashed lines represent a power law tail with exponent $-2$, as a guide to the eye.
\textbf{c}, The growth of new frontiers $S_n$ in real data feature a sublinear growth rate, lying in between the logarithmic and linear growth predicted by existing models.
\textbf{d}, We observe a consistent positive correlation between $W_n$ and $Q_n$ across all domains, suggesting recent record-breaking events are predictive of near-term occurrences.
\textbf{e}, The growth in variance $\langle S_n^2\rangle -\langle S_n\rangle ^2>>\langle S_n\rangle$ is systematically higher than expected, highlighting the long-term unpredictability for the number of record-breaking activities in these systems.}
  \label{fig:fig2}
\end{figure*}

\begin{figure*}[htbp]
    \centering
   \includegraphics[width=1\columnwidth]{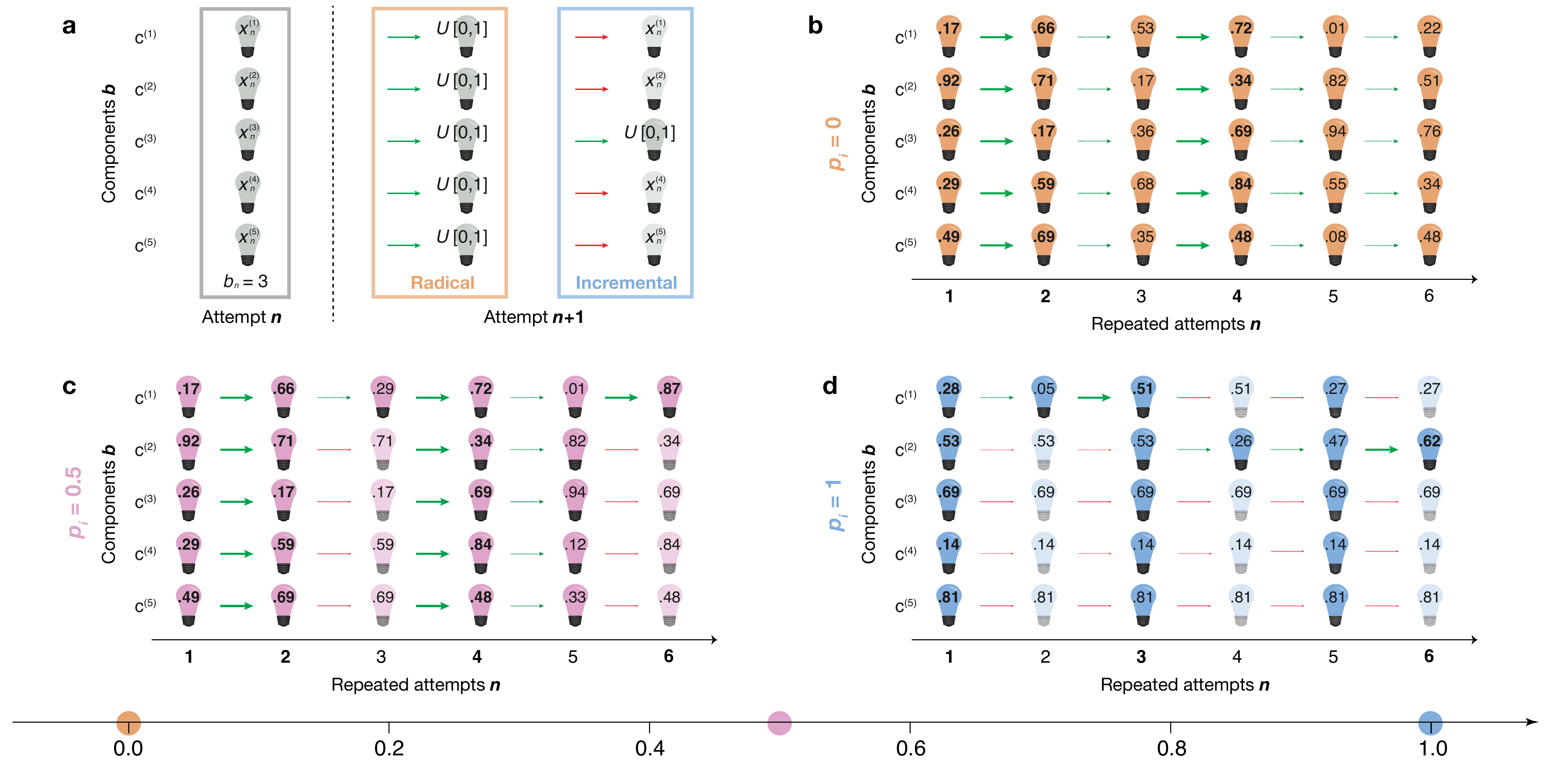}
  \caption{\textbf{Modeling incremental and radical innovations.} \textbf{a}, Each solution is viewed as a combination of components, where each component is associated with a score. The overall performance of the solution is a weighted sum of these components. To propose a new solution, an innovator can take either (i) a \emph{radical} approach with probability $p_r$ or (ii) an \emph{incremental} approach with probability $p_i=1-p_r$. In radical innovations, the innovator chooses to draw new random scores for \emph{all} components, independent of previous versions. While in incremental innovations, the innovator focuses on one component each time, replacing this component with a random draw while keeping all other components unchanged. \textbf{b-d}, illustrative examples of the model with $p_i=$ 0 (\textbf{b}), 0.5 (\textbf{c}), and 1 (\textbf{d}), respectively. Green arrows represent new random versions of a component. Attempts highlighted with boldface scores and arrows represent new frontier solutions.}
  \label{fig:fig3}
\end{figure*}

Despite its simplicity, the interplay introduced in $p$ model is sufficient to recover all empirical patterns documented in Fig.~\ref{fig:fig2}. By the $n$-th attempt, there are on average $Q_n^{(r)}\sim p_r n$ radical innovations and $Q_n^{(i)}\sim p_i Q_n$ incremental innovations (SI S3.2) on the current component of interest.
For any $0<p_r<1$, the waiting time until the next frontier $W_n$ follows,
\begin{equation}
   P(W_n\geq w)\approx \underbrace{\frac{Q_n^{(r)}}{wp_r+Q_n^{(r)}}}_{\text{Radical}}\underbrace{\frac{Q_n^{(i)}}{wp_i+Q_n^{(i)}}}_{\text{Incremental}} \approx \frac{1}{w/n+1}\frac{1}{w/Q_n+1}
   \label{eq:w_n_full0}
\end{equation}

\begin{figure*}[htbp]
    \centering
   \includegraphics[width=1\columnwidth]{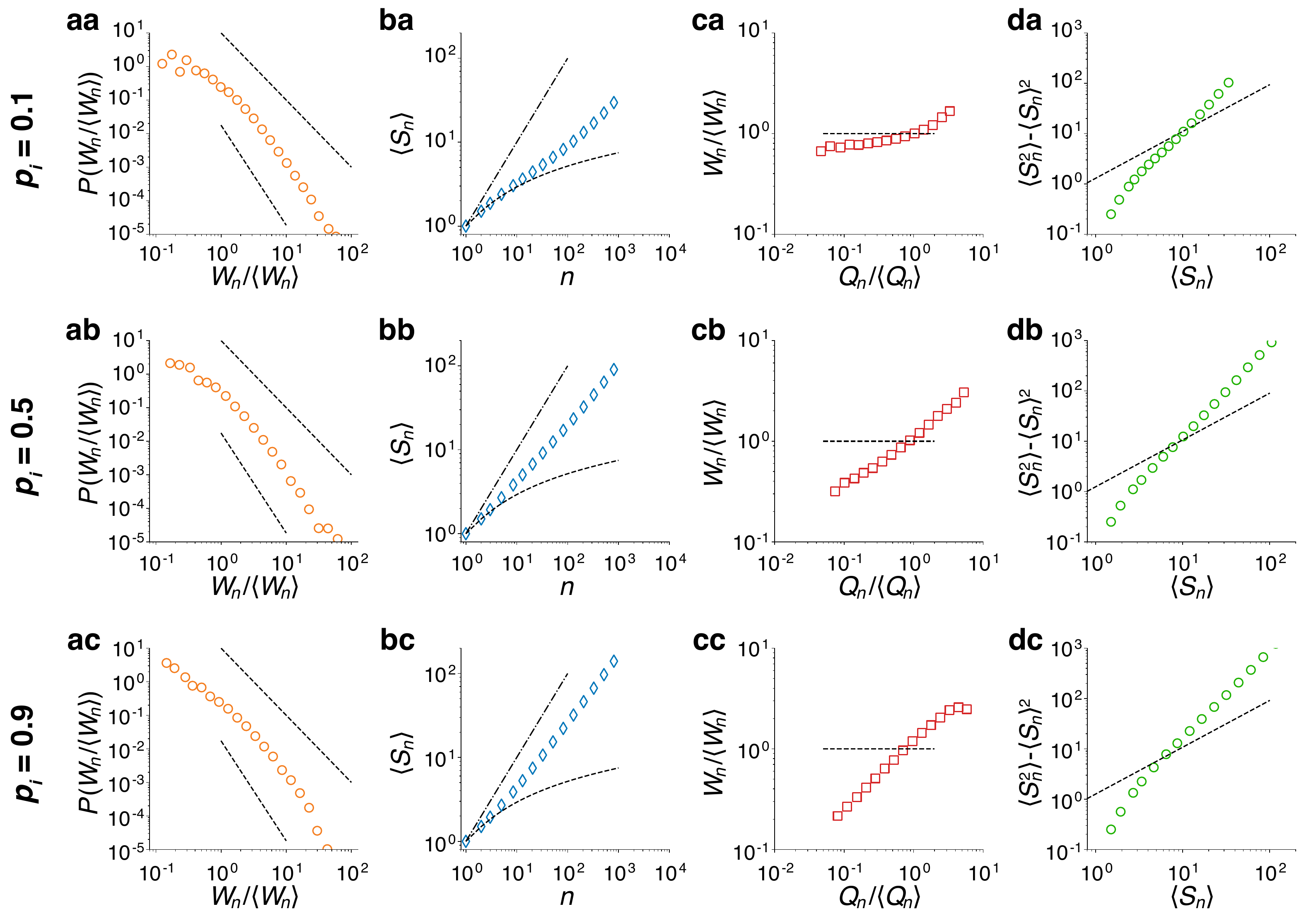}
  \caption{\textbf{Predictions of the $p$ model.} Despite its simplicity, the model systematically recovers all empirical patterns documented in Fig.~\ref{fig:fig2}. \textbf{a-d}, Same as Fig.~\ref{fig:fig2}\textbf{b-e}, but using model simulations with $p_i=$ 0.1 (\textbf{Row a}), 0.5 (\textbf{Row b}), and 0.9 (\textbf{Row c}), respectively.}
  \label{fig:fig4}
\end{figure*}

Eq.~(\ref{eq:w_n_full0}) uncovers a set of key insights, each corresponding to the empirical observations uncovered in Fig.~\ref{fig:fig2}.
First, note that $Q_n\sim \ln n$ (SI S3.1), a mean-field version of Eq.~(\ref{eq:w_n_full0}) writes
\begin{equation}
   P(W_n\geq w)\approx \biggl< \frac{1}{w/n+1}\frac{1}{w/Q_n+1}\biggr> \sim \frac{1}{w/n+1}\frac{1}{w/\ln n+1}
   \label{eq:w_n_full1}
\end{equation}
This provides an analytical explanation for the power law shape in Regularity 1: in the large $n$ limit (Fig.~\ref{fig:fig2}b,~\ref{fig:fig4}a): New improvements within short periods ($w\sim o(n)$) are dominated by incremental innovations, characterized by a power law decay of exponent $-2$ in the small-$w$ regime.
When $w$ grows to the scale of $O(n)$, however, both incremental and radical innovations play an important role in advancing new frontiers, leading to a truncated power law tail with exponent $-3$ (see SI S3 for more details):
\begin{equation}
   P(W_n=w)=-\frac{d}{dw}\left(\frac{1}{w/n+1}\frac{1}{w/\tilde{q}_n+1}\right)\sim
   \begin{cases}
    w^{-2}, &w\sim O(\ln n)\sim o(n)\\
    w^{-3}, &w\sim O(n)\\
    \end{cases}
\label{eq:prediction1}
\end{equation}

Second, Eq.~(\ref{eq:w_n_full1}) further indicates that the growth rate of new records is dominated by incremental innovations, which can be analytically solved by mapping it to a Levy flight (see SI S3 for more detail). Together, our analytical solution shows
\begin{equation}\langle S_n\rangle\sim p_i\frac{n}{\ln n}+(1-p_i)\ln n\sim \frac{n}{\ln n}.
\label{eq:prediction2}
\end{equation}
In other words, the growth rate falls between $\ln n$ predicted by Model A and $n$ predicted by Model B, consistent with Regularity 2 (Fig.~\ref{fig:fig2}c,~\ref{fig:fig4}b). It suggests a process where the chance of a new record decays slowly, because each new frontier creates a ``runway'' of incremental improvements that compound before saturating.

Third, Eq.~(\ref{eq:w_n_full0}) naturally connects $Q_n$ with $W_n$. In the small-$w$ regime, our model predicts that the ratio $t\equiv W_n/Q_n$ will converge to a limiting distribution
$f(t)\sim (t+1)^{-2}$,
consistent with the temporal auto-correlation in Regularity 3 (Fig.~\ref{fig:fig2}d,~\ref{fig:fig4}c). We further show that bursty incremental innovations lead to amplified variance in $S_n$, following
\begin{equation}\langle S_n^2\rangle-\langle S_n\rangle^2 \sim p_i^2\frac{n^2}{(\ln n)^3}+(1-p_i)^2\ln n\sim \frac{n^2}{(\ln n)^3}.
\label{eq:prediction3}
\end{equation}
These results not only explain the emergence of long-range correlations, but also allow us to derive the asymptotic growth rate of variance: $F_n\sim\langle S_n\rangle^2>>\langle S_n\rangle$, again consistent with our empirical observations (Fig.~\ref{fig:fig2}e,~\ref{fig:fig4}d).

Together, our analytical solutions show that the interplay between radical and incremental innovations naturally generates temporal clustering of breakthroughs. In the model, incremental innovation proceeds component by component: a successful improvement shifts the search to a new component, renewing the chance of further records and producing cascades that cluster in time. Conversely, when the current record is already high, the probability that either a local refinement or an occasional radical leap surpasses it becomes small, producing extended stasis. This state-dependent interplay yields all three empirical regularities at once: heavy-tailed waiting times, intermediate sublinear record growth, and positive temporal correlations.

Notably, while the model has one parameter $p$, all the key predictions across the three observations have $p$ canceled out (e.g. Eqs.~\ref{eq:prediction1}-\ref{eq:prediction3}).
This means, to the leading order, the $p$ model follows asymptotically the same behavior, independent of $p$. In other words, as long as one performs a non-zero fraction of incremental innovations, the leading-order predictions of the model hold the same for a large parameter space of $p_i$ (i.e. $0<p_i\leq 1$) (see SI S3 for detailed proof).
The $p$-model therefore identifies a universality class of frontier dynamics: in any system where occasional radical resets coexist with sustained incremental refinement around the frontier, all three key observations---heavy-tailed stasis, intermediate (sublinear) record accumulation, and temporal correlation with amplified variance---naturally arise, independent of domain-specific details. This explains why, despite the clear differences between the systems and domains we study, there is striking universality observed across these systems.

\section{Implications}

The $p$ model not only accounts for the universal regularities uncovered above, but also yields
falsifiable predictions about how structural features of the innovation environment reshape
the pace and clustering of progress. We highlight two such predictions, concerning the role of
knowledge disclosure and the asymmetric advantage of frontier access, test them empirically,
and then consider their broader consequences.

\textbf{Knowledge disclosure can accelerate frontier growth.} In our framework, open environments, where frontier solutions are visible to other participants, would directly amplify incremental innovations. Once a new frontier solution is disclosed, others can adopt it and test component-wise refinements, increasing the effective contribution of incremental moves and accelerating
the rate at which new records accumulate. By contrast, in closed environments where solutions are withheld,
most participants lack access to the current frontier. They must either refine their own (typically sub-frontier)
solutions or attempt larger leaps beyond an unseen state of the art, shifting progress toward a less incremental
and more leap-driven regime. The $p$ model therefore predicts that, all else equal, open systems should exhibit faster
accumulation of records than comparable closed systems.

\begin{figure*}[htbp]
    \centering
   \includegraphics[width=1\columnwidth]{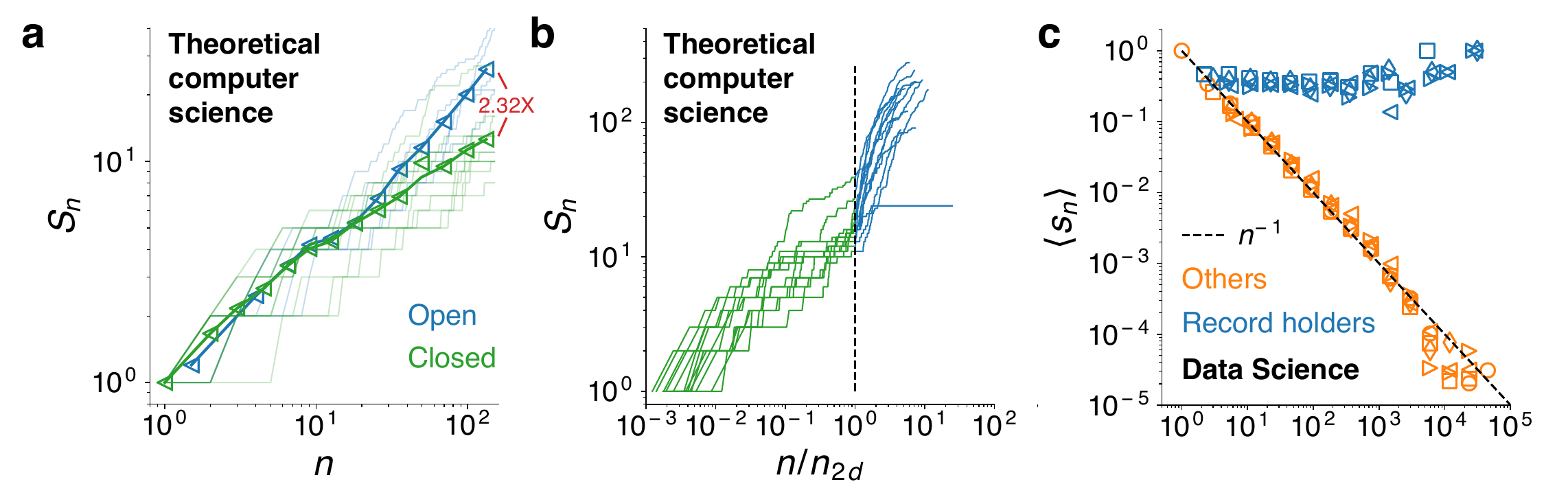}
  \caption{\textbf{Testing additional model predictions on empirical data.} \textbf{a}, In theoretical computer science competitions, fully open settings show a faster rate of record accumulation than non-disclosure intervals. \textbf{b}, In hybrid competitions, record accumulation accelerates after the transition from the closed phase to the open phase at the end of day 2.
\textbf{c}, In data-science competitions, current record holders maintain substantially higher record-setting rates ($s_n\sim 1/\ln n$) than other participants ($s_n\sim 1/n$).}
  \label{fig:fig5}
\end{figure*}

We test this prediction through a quasi-experimental setting: in the NP-hard algorithm competition series ($D_7$), earlier competitions operated in fully open mode, but later switched to a hybrid design in which
solutions are disclosed only after the first two days. Comparing the rate of record accumulation during the initial
non-disclosure phase of hybrid competitions with that of fully open competitions over matched intervals (Fig.~\ref{fig:fig5}a),
and comparing the open and non-disclosure phases within the same hybrid competitions (Fig.~\ref{fig:fig5}b), we find results are
consistent with the model's prediction: real-time access to frontier solutions amplifies incremental cascades and
accelerates frontier growth, with open phases producing approximately $2.3\times$ more record-setting events than
non-disclosure phases over the same interval.

\textbf{Frontier access confers a structural advantage.}
The model also makes predictions about who advances the frontier. In environments where frontier solutions are
not broadly disclosed, access to the current best solution is asymmetric. In such settings, those with direct access to
the frontier can apply incremental innovations to the state of the art, whereas others must surpass a frontier they
cannot directly observe. The $p$ model predicts that this asymmetry produces divergent record-breaking dynamics:
the record-breaking rate for frontier-access actors decays slowly (on the order of $1/\ln n$ in the model), whereas for
others it decays steeply (on the order of $1/n$), predicting a widening gap over time.

We test this prediction by separating attempts made by current record holders from those made by all other participants
in closed competition environments (Fig.~\ref{fig:fig5}c). Consistent with the model, record holders maintain a nearly constant
record-breaking rate over wide ranges of $n$, while the rate for other participants decays approximately as a power law
with exponent close to $-1$. This divergence quantifies the structural advantage conferred by direct access to the frontier solution
and illustrates how the effective balance between incremental refinement and larger leaps can shift systematically with
one's position relative to the state of the art.

\textbf{From theory to practice.}
These two predictions---that openness accelerates incremental cascades and that frontier access can confer a compounding
advantage---reflect general structural conditions that arise whenever innovators or organizations differ in their access to the current state
of the art. They therefore offer a mechanism-focused lens on contemporary debates about frontier-setting ecosystems.

Consider the development of frontier AI systems, one of the most consequential and contested innovation races underway
today. The landscape maps naturally onto the open-versus-closed structures studied here: some organizations release
architectures, code, or weights that enable broad incremental refinement, while others retain frontier solutions within
closed pipelines. Our framework predicts that this structural choice should have measurable dynamical consequences.
Open ecosystems should support faster accumulation of frontier-advancing contributions through cascades of incremental
refinements once a new advance is disclosed. In more closed settings, frontier holders retain the ability to iterate
incrementally on the best available model, while others face a steeper barrier to surpass an unseen frontier, making
progress more dependent on rarer leaps beyond the current state of the art. The divergence documented in Fig.~\ref{fig:fig5}c
provides a quantitative template for how differences in frontier access can translate into systematically different rates
of record-setting.

Beyond AI, similar access regimes arise in other frontier domains, including drug discovery, advanced semiconductor
manufacturing, and materials design, where control of frontier artifacts (data, tools, protocols, or evaluation infrastructure)
shapes who can refine the state of the art and how quickly improvements propagate.

More broadly, the $p$ model provides a minimal diagnostic framework for frontier-setting systems in which occasional radical resets coexist with sustained incremental refinement. By making punctuated progress measurable and comparable across domains, this framework opens a new quantitative agenda for understanding how frontiers advance. Taken together, our results indicate that punctuated progress may be governed by a small set of universal principles, with broad implications for how we understand, anticipate, and foster breakthroughs across science and technology.

\begin{acknowledgments}
We thank all members of the Center for Science of Science and Innovation (CSSI) for thoughtful discussions. This work is supported by the National Science Foundation 2404035, the Air Force Office of Scientific Research FA9550-19-1-0354, the Alfred P. Sloan Foundation G-2019-12485, and Schmidt Futures. The authors declare no competing financial interests.
\end{acknowledgments}

\bibliography{punctuated}

@book{kuhn1997structure,
  title={The structure of scientific revolutions},
  author={Kuhn, Thomas S},
  year={1962},
  publisher={University of Chicago Press}
}

@article{zhou2020power,
  title={Power-law distribution of degree--degree distance: A better representation of the scale-free property of complex networks},
  author={Zhou, Bin and Meng, Xiangyi and Stanley, H Eugene},
  journal={Proceedings of the national academy of sciences},
  volume={117},
  number={26},
  pages={14812--14818},
  year={2020},
  publisher={National Acad Sciences}
}

@article{duran2024multiscale,
  title={On the multiscale dynamics of punctuated evolution},
  author={Duran-Nebreda, Salva and Bentley, R Alexander and Vidiella, Blai and Spiridonov, Andrej and Eldredge, Niles and O’Brien, Michael J and Valverde, Sergi},
  journal={Trends in Ecology \& Evolution},
  year={2024},
  publisher={Elsevier}
}

@article{voitalov2019scale,
  title={Scale-free networks well done},
  author={Voitalov, Ivan and Van Der Hoorn, Pim and Van Der Hofstad, Remco and Krioukov, Dmitri},
  journal={Physical Review Research},
  volume={1},
  number={3},
  pages={033034},
  year={2019},
  publisher={APS}
}

@article{serafino2021true,
  title={True scale-free networks hidden by finite size effects},
  author={Serafino, Matteo and Cimini, Giulio and Maritan, Amos and Rinaldo, Andrea and Suweis, Samir and Banavar, Jayanth R and Caldarelli, Guido},
  journal={Proceedings of the National Academy of Sciences},
  volume={118},
  number={2},
  pages={e2013825118},
  year={2021},
  publisher={National Acad Sciences}
}

@article{liu2021understanding,
  title={Understanding the onset of hot streaks across artistic, cultural, and scientific careers},
  author={Liu, Lu and Dehmamy, Nima and Chown, Jillian and Giles, C Lee and Wang, Dashun},
  journal={Nature communications},
  volume={12},
  number={1},
  pages={5392},
  year={2021},
  publisher={Nature Publishing Group UK London}
}

@article{fleming2001recombinant,
  title={Recombinant uncertainty in technological search},
  author={Fleming, Lee},
  journal={Management science},
  volume={47},
  number={1},
  pages={117--132},
  year={2001},
  publisher={INFORMS}
}

@article{march1991exploration,
  title={Exploration and exploitation in organizational learning},
  author={March, James G},
  journal={Organization science},
  volume={2},
  number={1},
  pages={71--87},
  year={1991},
  publisher={INFORMS}
}

@article{kortum1997research,
  title={Research, patenting, and technological change},
  author={Kortum, Samuel S},
  journal={Econometrica: Journal of the Econometric Society},
  pages={1389--1419},
  year={1997},
  publisher={JSTOR}
}

@article{bak1993punctuated,
  title={Punctuated equilibrium and criticality in a simple model of evolution},
  author={Bak, Per and Sneppen, Kim},
  journal={Physical review letters},
  volume={71},
  number={24},
  pages={4083},
  year={1993},
  publisher={APS}
}

@article{gould1989punctuated,
  title={Punctuated equilibrium in fact and theory},
  author={Gould, Stephen Jay},
  journal={Journal of social and biological structures},
  volume={12},
  number={2-3},
  pages={117--136},
  year={1989},
  publisher={Elsevier}
}

@book{gould2009punctuated,
  title={Punctuated equilibrium},
  author={Gould, Stephen Jay and Gould, Stephen Jay},
  year={2009},
  publisher={Harvard University Press}
}

@article{gould1993punctuated,
  title={Punctuated equilibrium comes of age},
  author={Gould, Stephan Jay and Eldredge, Niles},
  journal={Nature},
  volume={366},
  number={6452},
  pages={223--227},
  year={1993},
  publisher={Nature Publishing Group UK London}
}

@article{loch1999punctuated,
  title={A punctuated-equilibrium model of technology diffusion},
  author={Loch, Christoph H and Huberman, Bernardo A},
  journal={Management Science},
  volume={45},
  number={2},
  pages={160--177},
  year={1999},
  publisher={INFORMS}
}

@article{valverde2015punctuated,
  title={Punctuated equilibrium in the large-scale evolution of programming languages},
  author={Valverde, Sergi and Sol{\'e}, Ricard V},
  journal={Journal of The Royal Society Interface},
  volume={12},
  number={107},
  pages={20150249},
  year={2015},
  publisher={The Royal Society}
}

@article{mokyr1990punctuated,
  title={Punctuated equilibria and technological progress},
  author={Mokyr, Joel},
  journal={The American Economic Review},
  volume={80},
  number={2},
  pages={350--354},
  year={1990},
  publisher={JSTOR}
}

@article{anderson1990technological,
  title={Technological discontinuities and dominant designs: A cyclical model of technological change},
  author={Anderson, Philip and Tushman, Michael L},
  journal={Administrative science quarterly},
  pages={604--633},
  year={1990},
  publisher={JSTOR}
}

@article{bettencourt2009scientific,
  title={Scientific discovery and topological transitions in collaboration networks},
  author={Bettencourt, Lu{\'\i}s MA and Kaiser, David I and Kaur, Jasleen},
  journal={Journal of Informetrics},
  volume={3},
  number={3},
  pages={210--221},
  year={2009},
  publisher={Elsevier}
}

@article{nevzorov1988records,
  title={Records},
  author={Nevzorov, Valerii Borisovich},
  journal={Theory of Probability \& Its Applications},
  volume={32},
  number={2},
  pages={201--228},
  year={1988},
  publisher={SIAM}
}

@article{krug2007records,
  title={Records in a changing world},
  author={Krug, Joachim},
  journal={Journal of Statistical Mechanics: Theory and Experiment},
  volume={2007},
  number={07},
  pages={P07001},
  year={2007},
  publisher={IOP Publishing}
}

@article{evenson1976stochastic,
  title={A stochastic model of applied research},
  author={Evenson, Robert E and Kislev, Yoav},
  journal={Journal of Political Economy},
  volume={84},
  number={2},
  pages={265--281},
  year={1976},
  publisher={The University of Chicago Press}
}

@article{terwiesch2008innovation,
  title={Innovation contests, open innovation, and multiagent problem solving},
  author={Terwiesch, Christian and Xu, Yi},
  journal={Management science},
  volume={54},
  number={9},
  pages={1529--1543},
  year={2008},
  publisher={INFORMS}
}

@article{henrich2004demography,
  title={Demography and cultural evolution: how adaptive cultural processes can produce maladaptive losses—the Tasmanian case},
  author={Henrich, Joseph},
  journal={American antiquity},
  volume={69},
  number={2},
  pages={197--214},
  year={2004},
  publisher={Cambridge University Press}
}

@article{kolodny2015evolution,
  title={Evolution in leaps: the punctuated accumulation and loss of cultural innovations},
  author={Kolodny, Oren and Creanza, Nicole and Feldman, Marcus W},
  journal={Proceedings of the National Academy of Sciences},
  volume={112},
  number={49},
  pages={E6762--E6769},
  year={2015},
  publisher={National Acad Sciences}
}

@article{rendell2010copy,
  title={Why copy others? Insights from the social learning strategies tournament},
  author={Rendell, Luke and Boyd, Robert and Cownden, Daniel and Enquist, Marquist and Eriksson, Kimmo and Feldman, Marc W and Fogarty, Laurel and Ghirlanda, Stefano and Lillicrap, Timothy and Laland, Kevin N},
  journal={Science},
  volume={328},
  number={5975},
  pages={208--213},
  year={2010},
  publisher={American Association for the Advancement of Science}
}

@article{miu2018innovation,
  title={Innovation and cumulative culture through tweaks and leaps in online programming contests},
  author={Miu, Elena and Gulley, Ned and Laland, Kevin N and Rendell, Luke},
  journal={Nature Communications},
  volume={9},
  number={1},
  pages={2321},
  year={2018},
  publisher={Nature Publishing Group UK London}
}

@article{clauset2009power,
  title={Power-law distributions in empirical data},
  author={Clauset, Aaron and Shalizi, Cosma Rohilla and Newman, Mark EJ},
  journal={SIAM review},
  volume={51},
  number={4},
  pages={661--703},
  year={2009},
  publisher={SIAM}
}

@article{tria2014dynamics,
  title={The dynamics of correlated novelties},
  author={Tria, Francesca and Loreto, Vittorio and Servedio, Vito Domenico Pietro and Strogatz, Steven H},
  journal={Scientific reports},
  volume={4},
  number={1},
  pages={5890},
  year={2014},
  publisher={Nature Publishing Group UK London}
}

@article{loreto2016dynamics,
  title={Dynamics on expanding spaces: modeling the emergence of novelties},
  author={Loreto, Vittorio and Servedio, Vito DP and Strogatz, Steven H and Tria, Francesca},
  journal={Creativity and universality in language},
  pages={59--83},
  year={2016},
  publisher={Springer}
}

@article{sole1996extinction,
  title={Extinction and self-organized criticality in a model of large-scale evolution},
  author={Sol{\'e}, Ricard V and Manrubia, Susanna C},
  journal={Physical Review E},
  volume={54},
  number={1},
  pages={R42},
  year={1996},
  publisher={APS}
}

@book{newman2003modeling,
  title={Modeling extinction},
  author={Newman, Mark EJ and Palmer, Richard G},
  year={2003},
  publisher={Oxford University Press, USA}
}

@article{kauffman1987towards,
  title={Towards a general theory of adaptive walks on rugged landscapes},
  author={Kauffman, Stuart and Levin, Simon},
  journal={Journal of theoretical Biology},
  volume={128},
  number={1},
  pages={11--45},
  year={1987},
  publisher={Elsevier}
}

@article{wiser2013long,
  title={Long-term dynamics of adaptation in asexual populations},
  author={Wiser, Michael J and Ribeck, Noah and Lenski, Richard E},
  journal={Science},
  volume={342},
  number={6164},
  pages={1364--1367},
  year={2013},
  publisher={American Association for the Advancement of Science}
}

@article{barabasi2005origin,
  title={The origin of bursts and heavy tails in human dynamics},
  author={Barabasi, Albert-Laszlo},
  journal={Nature},
  volume={435},
  number={7039},
  pages={207--211},
  year={2005},
  publisher={Nature Publishing Group UK London}
}

@article{margolin2013systematic,
  title={Systematic analysis of challenge-driven improvements in molecular prognostic models for breast cancer},
  author={Margolin, Adam A and Bilal, Erhan and Huang, Erich and Norman, Thea C and Ottestad, Lars and Mecham, Brigham H and Sauerwine, Ben and Kellen, Michael R and Mangravite, Lara M and Furia, Matthew D and others},
  journal={Science translational medicine},
  volume={5},
  number={181},
  pages={181re1--181re1},
  year={2013},
  publisher={American Association for the Advancement of Science}
}

@article{powell2009late,
  title={Late Pleistocene demography and the appearance of modern human behavior},
  author={Powell, Adam and Shennan, Stephen and Thomas, Mark G},
  journal={Science},
  volume={324},
  number={5932},
  pages={1298--1301},
  year={2009},
  publisher={American Association for the Advancement of Science}
}

@article{berdahl2017records,
  title={On the records},
  author={Berdahl, Andrew and Bhat, Uttam and Ferdinand, Vanessa and Garland, Joshua and Ghazi-Zahedi, Keyan and Grana, Justin and Grochow, Joshua A and Hobson, Elizabeth and Kallus, Yoav and Kempes, Christopher P and others},
  journal={arXiv preprint arXiv:1705.04353},
  year={2017}
}

@article{romanelli1994organizational,
  title={Organizational transformation as punctuated equilibrium: An empirical test},
  author={Romanelli, Elaine and Tushman, Michael L},
  journal={Academy of Management journal},
  volume={37},
  number={5},
  pages={1141--1166},
  year={1994},
  publisher={Academy of Management Briarcliff Manor, NY 10510}
}

@article{henderson1990architectural,
  title={Architectural innovation: The reconfiguration of existing product technologies and the failure of established firms},
  author={Henderson, Rebecca M and Clark, Kim B},
  journal={Administrative science quarterly},
  pages={9--30},
  year={1990},
  publisher={JSTOR}
}

@article{lemus2021dynamic,
  title={Dynamic tournament design: Evidence from prediction contests},
  author={Lemus, Jorge and Marshall, Guillermo},
  journal={Journal of Political Economy},
  volume={129},
  number={2},
  pages={383--420},
  year={2021},
  publisher={The University of Chicago Press Chicago, IL}
}

@article{elena1996punctuated,
  title={Punctuated evolution caused by selection of rare beneficial mutations},
  author={Elena, Santiago F and Cooper, Vaughn S and Lenski, Richard E},
  journal={Science},
  volume={272},
  number={5269},
  pages={1802--1804},
  year={1996},
  publisher={American Association for the Advancement of Science}
}

@article{sole1997self,
  title={Self-similarity of extinction statistics in the fossil record},
  author={Sole, Ricard V and Manrubia, Susanna C and Benton, Michael and Bak, Per},
  journal={Nature},
  volume={388},
  number={6644},
  pages={764--767},
  year={1997},
  publisher={Nature Publishing Group UK London}
}

@article{thompson2022complex,
  title={Complex cognitive algorithms preserved by selective social learning in experimental populations},
  author={Thompson, B and Van Opheusden, B and Sumers, T and Griffiths, TL},
  journal={Science},
  volume={376},
  number={6588},
  pages={95--98},
  year={2022},
  publisher={American Association for the Advancement of Science}
}

@article{baumgartner2009punctuated,
  title={Punctuated equilibrium in comparative perspective},
  author={Baumgartner, Frank R and Breunig, Christian and Green-Pedersen, Christoffer and Jones, Bryan D and Mortensen, Peter B and Nuytemans, Michiel and Walgrave, Stefaan},
  journal={American Journal of Political Science},
  volume={53},
  number={3},
  pages={603--620},
  year={2009},
  publisher={Wiley Online Library}
}

@article{stuart1996local,
  title={Local search and the evolution of technological capabilities},
  author={Stuart, Toby E and Podolny, Joel M},
  journal={Strategic management journal},
  volume={17},
  number={S1},
  pages={21--38},
  year={1996},
  publisher={Wiley Online Library}
}

@book{christensen2013innovator,
  title={The innovator's dilemma: when new technologies cause great firms to fail},
  author={Christensen, Clayton M},
  year={2013},
  publisher={Harvard Business Review Press}
}

@article{bednorz1986possible,
  title={Possible high T c superconductivity in the Ba- La- Cu- O system},
  author={Bednorz, J George and M{\"u}ller, K Alex},
  journal={Zeitschrift f{\"u}r Physik B Condensed Matter},
  volume={64},
  number={2},
  pages={189--193},
  year={1986},
  publisher={Springer}
}

@article{song2010modelling,
  title={Modelling the scaling properties of human mobility},
  author={Song, Chaoming and Koren, Tal and Wang, Pu and Barab{\'a}si, Albert-L{\'a}szl{\'o}},
  journal={Nature physics},
  volume={6},
  number={10},
  pages={818--823},
  year={2010},
  publisher={Nature Publishing Group UK London}
}

@article{koren2007leapover,
  title={Leapover lengths and first passage time statistics for L{\'e}vy flights},
  author={Koren, Tal and Lomholt, Michael A and Chechkin, Aleksei V and Klafter, Joseph and Metzler, Ralf},
  journal={Physical review letters},
  volume={99},
  number={16},
  pages={160602},
  year={2007},
  publisher={APS}
}

@article{zaburdaev2015levy,
  title={L{\'e}vy walks},
  author={Zaburdaev, Vasily and Denisov, Sergey and Klafter, Joseph},
  journal={Reviews of Modern Physics},
  volume={87},
  number={2},
  pages={483},
  year={2015},
  publisher={APS}
}

@article{atkinson2008languages,
  title={Languages evolve in punctuational bursts},
  author={Atkinson, Quentin D and Meade, Andrew and Venditti, Chris and Greenhill, Simon J and Pagel, Mark},
  journal={Science},
  volume={319},
  number={5863},
  pages={588--588},
  year={2008},
  publisher={American Association for the Advancement of Science}
}

@article{gersick1991revolutionary,
  title={Revolutionary change theories: A multilevel exploration of the punctuated equilibrium paradigm},
  author={Gersick, Connie JG},
  journal={Academy of management review},
  volume={16},
  number={1},
  pages={10--36},
  year={1991},
  publisher={Academy of Management Briarcliff Manor, NY 10510}
}

@article{pagel2006large,
  title={Large punctuational contribution of speciation to evolutionary divergence at the molecular level},
  author={Pagel, Mark and Venditti, Chris and Meade, Andrew},
  journal={Science},
  volume={314},
  number={5796},
  pages={119--121},
  year={2006},
  publisher={American Association for the Advancement of Science}
}

@article{thurner2010schumpeterian,
  title={Schumpeterian economic dynamics as a quantifiable model of evolution},
  author={Thurner, Stefan and Klimek, Peter and Hanel, Rudolf},
  journal={New Journal of Physics},
  volume={12},
  number={7},
  pages={075029},
  year={2010},
  publisher={IOP Publishing}
}

@article{ott2022mapping,
  title={Mapping global dynamics of benchmark creation and saturation in artificial intelligence},
  author={Ott, Simon and Barbosa-Silva, Adriano and Blagec, Kathrin and Brauner, Jan and Samwald, Matthias},
  journal={Nature Communications},
  volume={13},
  number={1},
  pages={6793},
  year={2022},
  publisher={Nature Publishing Group UK London}
}

@article{martinez2021research,
  title={Research community dynamics behind popular AI benchmarks},
  author={Mart{\'\i}nez-Plumed, Fernando and Barredo, Pablo and Heigeartaigh, Sean O and Hern{\'a}ndez-Orallo, Jos{\'e}},
  journal={Nature Machine Intelligence},
  volume={3},
  number={7},
  pages={581--589},
  year={2021},
  publisher={Nature Publishing Group UK London}
}

@article{osiurak2021technical,
  title={Technical reasoning is important for cumulative technological culture},
  author={Osiurak, Fran{\c{c}}ois and Lasserre, Salom{\'e} and Arbanti, Julie and Brogniart, Jo{\"e}l and Bluet, Alexandre and Navarro, Jordan and Reynaud, Emanuelle},
  journal={Nature Human Behaviour},
  volume={5},
  number={12},
  pages={1643--1651},
  year={2021},
  publisher={Nature Publishing Group UK London}
}

@article{karsai2012universal,
  title={Universal features of correlated bursty behaviour},
  author={Karsai, M{\'a}rton and Kaski, Kimmo and Barab{\'a}si, Albert-L{\'a}szl{\'o} and Kert{\'e}sz, J{\'a}nos},
  journal={Scientific reports},
  volume={2},
  number={1},
  pages={397},
  year={2012},
  publisher={Nature Publishing Group UK London}
}

@article{derex2019causal,
  title={Causal understanding is not necessary for the improvement of culturally evolving technology},
  author={Derex, Maxime and Bonnefon, Jean-Fran{\c{c}}ois and Boyd, Robert and Mesoudi, Alex},
  journal={Nature human behaviour},
  volume={3},
  number={5},
  pages={446--452},
  year={2019},
  publisher={Nature Publishing Group UK London}
}
\clearpage

\begin{table}[htbp]
\rowcolors{2}{lightgray}{darkgray}
\begin{tabular}{cc|c|c|c}
\hline
\textbf{~} & \textbf{Domain} & \textbf{Data source}& \textbf{Tasks} & \textbf{Attempts} \\
\hline
$D_1$& Superconducting Materials  & Reaxys &  33 &38,576 \\
$D_2$& Structural Biology & Protein Data Bank &  3,644 &143,343 \\
$D_3$& Artificial Intelligence & Paperswithcode &  385 &10,439 \\
$D_4$& Computational Biomedicine & DREAM &  70 &15,529 \\
\hline
$D_5$& Data science & Kaggle & 2,067 &5,792,702  \\
$D_6$& Data Science & TopCoder &  414 &188,807 \\
$D_7$& Theoretical Computer Science & Matlab &  19 & 45,914 \\
\hline
$D_8$& Technology &  Formula 1 &  41 &589,081 \\
$D_9$& Technology &  Experiments &  56 &1,400 \\
\hline
\end{tabular}
\caption{Descriptions of datasets ($D_1-D_9$) used in this paper.}
\label{Tab1}
\end{table}
\clearpage

\clearpage

\clearpage

\clearpage

\clearpage

\clearpage

\end{document}